\documentclass[12pt]{article}
\usepackage{cite}
\usepackage{graphicx}
\pdfoutput=1

\newcommand{\Refs}[1]{Refs.~\cite{#1}}
\newcommand{\Ref}[1]{Ref.~\cite{#1}}
\newcommand{\Eq}[1]{equation~(\ref{#1})}
\newcommand{\Fig}[1]{figure~\ref{#1}}
\newcommand{\Figab}[1]{figures~\ref{#1}(a) and~\ref{#1}(b)}
\newcommand{\VI}{\mbox{$V-I$}}
\newcommand{\Lsim}{\stackrel{<}{_\sim}}
\newcommand{\Gsim}{\stackrel{>}{_\sim}}
\newcommand{\kB}{\mbox{$k_{\rm B}$}}
\newcommand{\electron}{\mbox{$e$}}
\newcommand{\Tc}{\mbox{$T_c$}}
\newcommand{\Tco}{\mbox{$T_{c0}$}}
\newcommand{\Tccero}{\mbox{$T_{c0}$}}
\newcommand{\Tcceromedia}{\mbox{$\overline{T}_{c0}$}}
\newcommand{\TBKTmedia}{\mbox{$\overline{T}_{\rm BKT}$}}
\newcommand{\TKTmedia}{\TBKTmedia}
\newcommand{\TBKT}{\mbox{$T_{\rm BKT}$}}
\newcommand{\TKT}{\TBKT}
\newcommand{\deltaBKT}{\mbox{$\Delta_{\rm BKT}$}}
\newcommand{\deltaKT}{\deltaBKT}

\newcommand{\deltaTccero}{\mbox{$\Delta T_{c0}$}}
\newcommand{\jiab}{\mbox{$\xi_{ab}$}}

\newcommand{\ie}{\mbox{\it i.e.}}
\newcommand{\eg}{\mbox{\it e.g.}}

\newcommand{\sigman}{\mbox{$\sigma_{\rm n}$}}
\newcommand{\Ds}{\mbox{$\Delta\sigma$}}
\newcommand{\xiabo}{\mbox{$\xi_{ab}(0)$}}
\newcommand{\xiabKT}{\mbox{$\xi_{ab{\rm BKT}}$}}

\newcommand{\YBCO}{\mbox{${\rm Y}{\rm Ba}_2{\rm Cu}_3{\rm O}_{7-\delta}$}}
\newcommand{\TlBCCO}{\mbox{${\rm Tl}_2{\rm Ba}_2{\rm Ca}{\rm Cu}_2{\rm O}_8$}}

\begin{document}

\newcommand{\titulo}{Effects of critical temperature inhomogeneities on the voltage-current characteristics of a planar superconductor near the Berezinskii-Kosterlitz-Thouless transition}

\newcommand{\autor}{N.~Cot\'on, M.V.~Ramallo and F.~Vidal}

\newcommand{\direccion}{Laboratorio de Baixas Temperaturas e Supercondutividade LBTS,\\ Departamento de F\'{\i}sica da Materia Condensada,\\ Universidade de Santiago de Compostela, E-15782 Santiago de Compostela, Spain}

\begin{center}
  \Large\bf
\titulo\\  \end{center}\mbox{}\vspace{-1cm}\\ 

\begin{center}\normalsize\autor\end{center} 

\begin{center}\normalsize\it\direccion\end{center}

%%%%---abstract---

\mbox{}\vskip0.5cm{\bf Abstract. }
We analyze numerically how the voltage-current (\VI) characteristics near the so-called Berezinskii-Kosterlitz-Thouless (BKT) transition of 2D superconductors are affected by a random spatial Gaussian distribution of critical temperature inhomogeneities with long characteristic lengths (much larger than the in-plane superconducting coherence length amplitude). Our simulations allow to quantify the broadening around the average BKT transition temperature \TBKTmedia\ of both the exponent $\alpha$ in $V\propto I^{\alpha}$ and of the resistance $V/I$. These calculations reveal that strong spatial redistributions of the local current will occur around the transition as either $I$ or the temperature $T$ are varied. Our results also support that the condition $\alpha=3$ provides a good estimate for the location of the average BKT transition temperature \TBKTmedia, and that extrapolating to $\alpha\rightarrow 1$ the $\alpha(T)$ behaviour well below the transition provides a good estimate for the average mean-field critical temperature \Tcceromedia.\vspace{4cm}

\mbox{}\hfill{\footnotesize {\tt ramallo@cond-mat.eu}}
\thispagestyle{empty}

\newpage
\setlength{\baselineskip}{18pt}

%%%%---main body of the article

\newpage
\setlength{\baselineskip}{18pt}

%%%%---main body of the article

\section{Introduction}

As first suggested by Berezinskii\cite{Berezinskii} and by Kosterlitz and Thouless\cite{KosterlitzThouless} (BKT) for superfluids, and later for superconductors by various authors (see, \eg, \Refs{Beasley,Doniach,HalperinNelson}), one of the main features to be expected in the normal-superconducting transition of a two-dimensional (2D) and homogeneous type-II superconductor is the appearance at zero applied magnetic field of two critical temperatures: Namely, the transition is split into {\it i)}~the mean-field critical temperature \Tccero, where it first becomes favorable (in terms of free-energy optimization) to form Cooper pairs and vortices, and  {\it ii)}~the so-called BKT critical temperature $\TBKT<\Tccero$, where it first becomes favorable that vortices of opposite fluxoid quantization bind into pairs (vortex-antivortex pairs). Among the experimental features revealing this splitting of the transition, probably the most significant is the appearance of a strongly non-Ohmic behaviour in the voltage-current \VI\  characteristics below \TBKT, with the exponent $\alpha$ in $V\propto I^\alpha$ jumping at \TBKT\ itself to the value $\alpha=3$.\cite{HalperinNelson} This feature has been, in fact, commonly used to experimentally demonstrate the existence of a BKT transition, both in low-\Tccero\ 2D structures (see, \eg, \Refs{LTSC-1,LTSC-2,LTSC-3,Science,Nature}) and in high-\Tccero\ superconducting cuprates (HTSC) (see, \eg, \Refs{HTSC-1,PRB40VI,PRB42,HTSC-3,HTSC-4}). We note that in the case of the HTSC materials even bulk samples are expected to undergo a BKT transition, due to the anisotropic 2D-like layered structure of these superconductors.\cite{nota} In fact, in HTSC with optimal doping the measurements suggest that the difference $\Tccero-\TBKT$, henceforth noted as \deltaBKT, may be rather large, of about 2K.\cite{HTSC-1,PRB40VI,PRB42,HTSC-3,HTSC-4} This value agrees well with the theoretical predictions, as \deltaBKT\ may be approximated by the so-called Levanyuk-Ginzburg criterion \cite{eGL} that also estimates the size of the full-critical (or non-Gaussian) region of superconductivity fluctuations above \Tccero; this full-critical region has been determined on both theoretical and experimental grounds to span over about 2K in optimally-doped HTSC.\cite{eGL,MosqueiraRamallo,capitulolibro}

In spite of the fact that most real samples are expected to have some degree of inhomogeneities of the values of \Tccero\ (and of \TBKT), almost no calculations have been done on how a spatially-random distribution of \Tccero-values will affect the BKT non-Ohmic characteristics. Recently, Benfatto and coworkers\cite{Benfatto}  proposed a renormalization-group study for such situation, but their approach implies to estimate the global resistivity of the sample just averaging the ones of the homogeneous domains. This assumption could be expected to be adequate only if the current itself is homogeneous in the sample (see also below).   In this paper, we use mesh-circuit numerical analyses to study the BKT non-Ohmic features that result from considering a 2D type-II superconductor having a Gaussian distribution of inhomogeneities of \Tccero\ and \TBKT, randomly located in space and with long characteristic lengths (much larger, in particular, than the in-plane superconducting coherence length amplitude). Our analysis allows to obtain the evolution of $\alpha$ with temperature, and also shows that significant current redistributions occur in the sample as $T$ and $I$ are varied. We believe that our results are applicable to measurements (\eg, those in \Refs{HTSC-1,PRB40VI,PRB42,HTSC-3,HTSC-4}) that show an $\alpha(T)$ jump near \TBKT\ well smoother than predicted by the theory of homogeneous superconductors.

To get a first glimpse of some of the main difficulties of the proposed problem, let us consider in this preface two oversimplified cases of \Tc-inhomogeneities: {\it i)}~First, a film with domains corresponding to rectangular halves situated with respect to the current contacts in series configuration, and {\it ii)}~the same situation but with domains in parallel with respect to the current contacts. In the first case, obviously the current passes through both zones without any spatial redistribution as $T$ or $I$ varies, and the total resistivity is simply the average of the resistivities of both zones. However, even in this uncomplicated case the exponent $\alpha$ of the whole sample will not be just the average of the $\alpha$-values of both zones, as the larger contribution to the total $V$ drop (and hence to the global $V\propto I^{\alpha}$ behaviour) happens in the zone with {\it larger} resistivity. Now we consider the situation {\it ii)} where the two \Tc-zones are in parallel configuration. In this case, when either $T$ or $I$ are varied, and with them the quotient of the resistivities of the two zones, there will be spatial redistributions of the currents. In fact, at some $T$-$I$ combinations, these redistributions will be extreme enough as to become percolating-like. Correspondingly, the contribution to $\alpha$ from both zones will be now very different to the one in case {\it i)}. For instance, the global resistivity will greatly differ from the average of both zones and the main contribution to $\alpha$ will be given now by the zone with {\it lower} resistivity. Obviously, a realistic model of a randomly inhomogeneous sample will include many domains both in series and in parallel. It will be then nontrivial to know if $\alpha$ will be dominated by the zones with higher or lower \Tc\ values. Also, a sizeable part of the $T,I$ phase diagram will be affected by percolating-like effects. This makes it difficult to successfully formulate in a comprehensive $T,I$ range an analytic estimate for $\alpha$  in terms of simple averages or even of effective-medium approaches. \cite{MazaVidalPomarRamallo} Because of these difficulties, we have chosen numerical simulation methods to analyze the problem.

This paper is organized as follows: In section 2 we will briefly summarize the theory expressions for the \VI\  curves of an homogeneous superconducting film near its BKT transition. In section 3 we will detail our numerical algorithms and procedures. In section 4 we will present and discuss the resulting \VI\  curves, and also briefly compare them with some previous experimental measurements by other authors, focusing on the case of HTSC.\cite{PRB40VI,PRB42} Section~5 summarizes our conclusions.

\mbox{}

\section{Summary of the theoretical approaches for homogeneous systems}\label{sec:ii}

\subsection{Superconducting contributions to the conductivity near the BKT transition at zero applied magnetic field} 

To study the \VI\  characteristics of a 2D type-II superconductor with a random spatial Gaussian distribution of \TBKT\ and \Tccero, we will use as starting point the \VI\ expressions for homogeneous superconductors as proposed by Halperin and Nelson (HN) in \Ref{HalperinNelson}. To summarize these expressions in a way convenient for our purposes, we consider the different $T$-regimes that appear as $T$ moves from higher to lower values.

{\it i) Temperatures $T>\TBKT$.} In this $T$-range the superconducting contribution to the electrical conductivity corresponds to the existence of thermal fluctuations of the  order parameter without vortex-antivortex binding effects, which may be taken into account by the Gaussian-Ginzburg-Landau approach (appropriate for $T\Gsim\Tccero+\deltaBKT$)  and  the $XY$-model renormalization-group approach (for $\TBKT<T\Lsim\Tccero+\deltaBKT$). HN have proposed\cite{HalperinNelson} an useful interpolation formula that covers the results of both of these approaches, with accuracy well sufficient for our present purposes:
%%%%%%%%%%%%%%%%%%%%%%%%%%%%%%%%%%%%%%%%%%%%%%%%%
\begin{equation} \label{ecuacionHalperinNelson}
\Ds=\frac{0.37\sigma_n}{b_0}\,{\rm sinh^2}\sqrt{\frac{b_0\, \deltaBKT}{T-\TBKT}}   \;\;\;\;\mbox{(for}\;T>\TBKT\mbox{),}
\end{equation}
%%%%%%%%%%%%%%%%%%%%%%%%%%%%%%%%%%%%%%%%%%%%%%%%%
where \Ds\ and $\sigma_n$ are, respectively, the superconducting and normal contributions to the in-plane electrical conductivity,  and $b_0$ is a dimensionless parameter for which HN do not propose any definite value, stating only that it may be expected to be of the order of unity. However, let us already mention here that the value of $b_0$ will be further constrained when considering the expressions for  $T<\TBKT$ (see point {\it ii} below). Note also that \Eq{ecuacionHalperinNelson} corresponds to an Ohmic \Ds\ (if considering an Ohmic $\sigma_n$, as will be done in all this work). We also note that we have checked that using the more accurate expressions \cite{Beasley,Doniach,HalperinNelson,RamalloPomarVidal} for \Ds\ instead of the interpolated \Eq{ecuacionHalperinNelson} does not affect in any significant way the main results presented in this paper.

{\it ii) Temperatures $T<\TBKT$.} In this $T$-range, the relevant degrees of freedom for the superconducting fluctuations are vortices and antivortices. Again,  this region may be divided into two, according to the strength of those fluctuations:  In the range of temperatures closer to \TBKT\ than \deltaBKT\ the fluctuations are full-critical, while for lower temperatures the superconductor follows a conventional Ginzburg-Landau behaviour. Following again the ideas of HN we summarize into one common expression the \Ds\  results for both $T$-regions at  fixed current density $j$:\cite{LTSC-1}
%%%%%%%%%%%%%%%%%%%%%%%%%%%%%%%%%%%%%%%%%%%%%%%%%
\begin{equation} \label{ecuaciondebajoTBKT-Ds}
\Ds=\frac{\sigma_n}{2(\alpha-3)}\,\left(\frac{j}{j_0}\right)^{1-\alpha}   \;\;\;\;{\rm(for}\;T<\TBKT{\rm ),}
\end{equation}
%%%%%%%%%%%%%%%%%%%%%%%%%%%%%%%%%%%%%%%%%%%%%%%%%
where $j_0=\electron\kB\TKT/(\hbar d \xiabKT)$ is the Ginzburg-Landau critical current density at \TKT, \electron\ the electron charge, \kB\ the Boltzmann constant, $\hbar$ the reduced Planck constant, $d$ the sample thickness, $\xiabKT=\xiabo(\Tco/\deltaKT)^{1/2}$ and \xiabo\ the Ginzburg-Landau in-plane coherence length extrapolated to respectively \TBKT\ and $T=0$K, $\alpha=\mbox{max}(\alpha_T,\alpha_J)$, $\alpha_J=3-1/\ln(j/j_0)^2$, and $\alpha_T$ is:
%%%%%%%%%%%%%%%%%%%%%%%%%%%%%%%%%%%%%%%%%%%%%%%%%
\begin{equation} \label{ecuaciondebajoTBKT-x}
\alpha_T=\left\{
\begin{array}{ll}
3+\pi\sqrt{\frac{\TBKT-\mbox{$T$}}{\mbox{$b_0$}\,\deltaBKT}} &\mbox{(for $\TBKT-\deltaBKT<T<\TBKT$),}\\
&\\
1+2b_1\frac{\Tco-\mbox{$T$}}{\deltaBKT} &\mbox{(for $T<\TBKT-\deltaBKT$).}\end{array}
\right.
\end{equation}
%%%%%%%%%%%%%%%%%%%%%%%%%%%%%%%%%%%%%%%%%%%%%%%%%
Note that $\alpha_T\rightarrow3$ when $T\rightarrow\TBKT$ from below, and then at \TBKT\ it is $\Ds\propto j^{-2}$, \ie, $V\propto I^3$ (neglecting the small effect of $\alpha_J$ and of any other contributions to the conductivity). The parameter $b_1$ in \Eq{ecuaciondebajoTBKT-x} takes into account the variations of the value of \deltaBKT\ with respect  to the purely 2D non-fluctuation GL value.\cite{nota}  In the theoretical simulations done in this work, for simplicity we will neglect such differences and take $b_1=1$ (except for the comparison with the experimental data shown in \Fig{Fig6} where $b_1$ has been fine-tuned to the value $b_1=2$). Note also that to ensure continuity at $T=\TBKT-\deltaBKT$ of \Eq{ecuaciondebajoTBKT-x} (and thus of \Eq{ecuaciondebajoTBKT-Ds})  it is needed that the parameter $b_0$ takes the value $b_0=\pi^2/(4b_1-2)^2$, rather than being a somewhat free choice as suggested by HN.\cite{HalperinNelson}  Finally, note also that the \VI\ exponent $\alpha$ resulting from \Eq{ecuaciondebajoTBKT-x} at $T<\TBKT-\deltaBKT$ {\it extrapolates} to the Ohmic value at $T=\Tccero$, suggesting a possible procedure for the experimental identification of the mean-field critical temperature.

\subsection{Other contributions to the conductivity near the BKT transition}

To obtain the total conductivity $\sigma$ of the superconductor we must add to the above formulas for \Ds\ the contributions from the rest of electrical transport channels in the system, mainly the conductivity $\sigma_n$ due to the normal-state carriers:
%%%%%%%%%%%%%%%%%%%%%%%%%%%%%%%%%%%%%%%%%%%%%%%%%
\begin{equation} \label{sigmatotal}
\sigma=\Ds+\sigma_n.
\end{equation}
%%%%%%%%%%%%%%%%%%%%%%%%%%%%%%%%%%%%%%%%%%%%%%%%%
Indeed $\sigma_n$ will be negligible against \Ds\ for  $T\Lsim\TKT$, but for larger temperatures it won't be so. Note also that $\sigma_n$ may be $T$-dependent. For instance in the optimally-doped HTSC it may be well approximated as inversely proportional to $T$ (as will in fact be used in our simulations, see below). We will always consider in this paper that $\sigma_n$ itself is Ohmic.

Other contributions to the total conductivity will be neglected in this paper, but we note here that in certain specific experimental circumstances they could become appreciable: For instance, for $T\Lsim\TKT$ we neglect the Ohmic conductivity that may appear at very low intensities  when the vortex-antivortex pair breaking processes involve distances larger than the inhomogeneity size (see, \eg, \Ref{HalperinNelson}).  We also neglect the non-Ohmic contributions to \Ds\ that may appear above \TKT\ due to the superconducting fluctuations  not related to the vortex-antivortex correlations. The latter contributions have been thoroughly explored previously (see, \eg, \Refs{Hurault,Schmid,Varlamov,Carlos}) and today are well known to be significative only at electrical fields much larger than those considered in our study (see, \eg, figure~(1) of \Ref{PuicaLang}). Finally, we also mention that we neglect the indirect contributions to the paraconductivity above the transition, such as the so-called Maki-Thomson and density-of-states contributions.\cite{capitulolibro} It is today quite well accepted that such contributions are negligible in the case of HTSC\cite{capitulolibro,RamalloPomarVidal,PomarDiazPomarRamallo} although the situation is not as clear in the case of low-\Tc\ superconducting thin films.\cite{Patton} We have checked however that including any of such contributions in our simulations does not  qualitatively affect the main results presented in this paper.

\mbox{}

\section{Procedure for the numerical simulation of the \VI\  characteristics near the BKT transition of an inhomogeneous superconductor}

\begin{figure}[b!]
\begin{center}\mbox{}\vspace{1.5cm}\\
\includegraphics[width=1\textwidth]{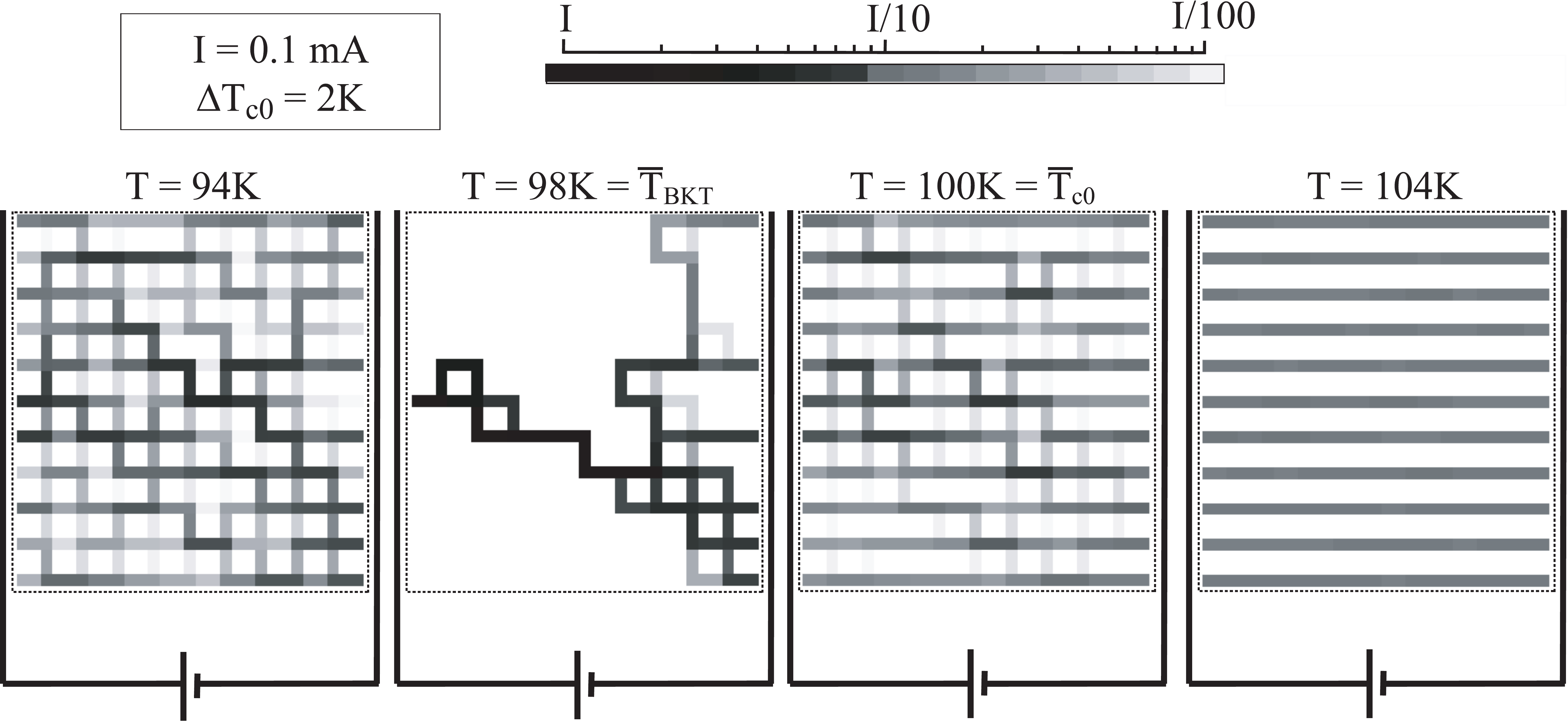}\end{center}
\caption{Some examples of the evolution of the current distributions obtained in our simulations when the temperature is varied and the bias intensity is kept constant. The simulation sample is composed by a $10\times10$ mesh of resistors corresponding to a random spatial Gaussian distribution of mean-field critical temperatures with average \Tcceromedia\ and full-width at half-maximum \deltaTccero, with the value of $\deltaBKT=\Tccero-\TBKT$ being the same for all resistors. In the pictured example, $\Tcceromedia=100$K, $\deltaBKT=2$K (so that $\TBKTmedia=98$K), $\deltaTccero=2$K, the sample width and length is $w=10^{-3}{\rm m}$ and its thickness is $d=100{\rm nm}$.  We also used (see main text) $b_1=1$, $\sigman=10^8T^{-1}$K/$\Omega$m and $\jiab(0)=1$nm.  The bias current $I$ is the one provided by the external current source, which is connected to opposite borders of the superconductor with zero-resistance contacts. Note the strong spatial redistribution of currents inside the superconductor as $T$ is varied through the transition, indicating that at each temperature different regions of the superconductor determine the global resistivity, and its Ohmic or non-Ohmic character. These redistributions are specially intense near the average BKT transition temperature where the current flows through only a few less resistive paths. It may be also observed that for $T\gg\Tcceromedia$ almost all of the current flows longitudinal and uniformly, while for $T<\TBKTmedia$ transversal current paths are significant.}
\label{Fig1} 
\end{figure}

\begin{figure}[b!]
\mbox{}\vspace{1.0cm}\\
\mbox{}\hfill\includegraphics[width=0.95\textwidth]{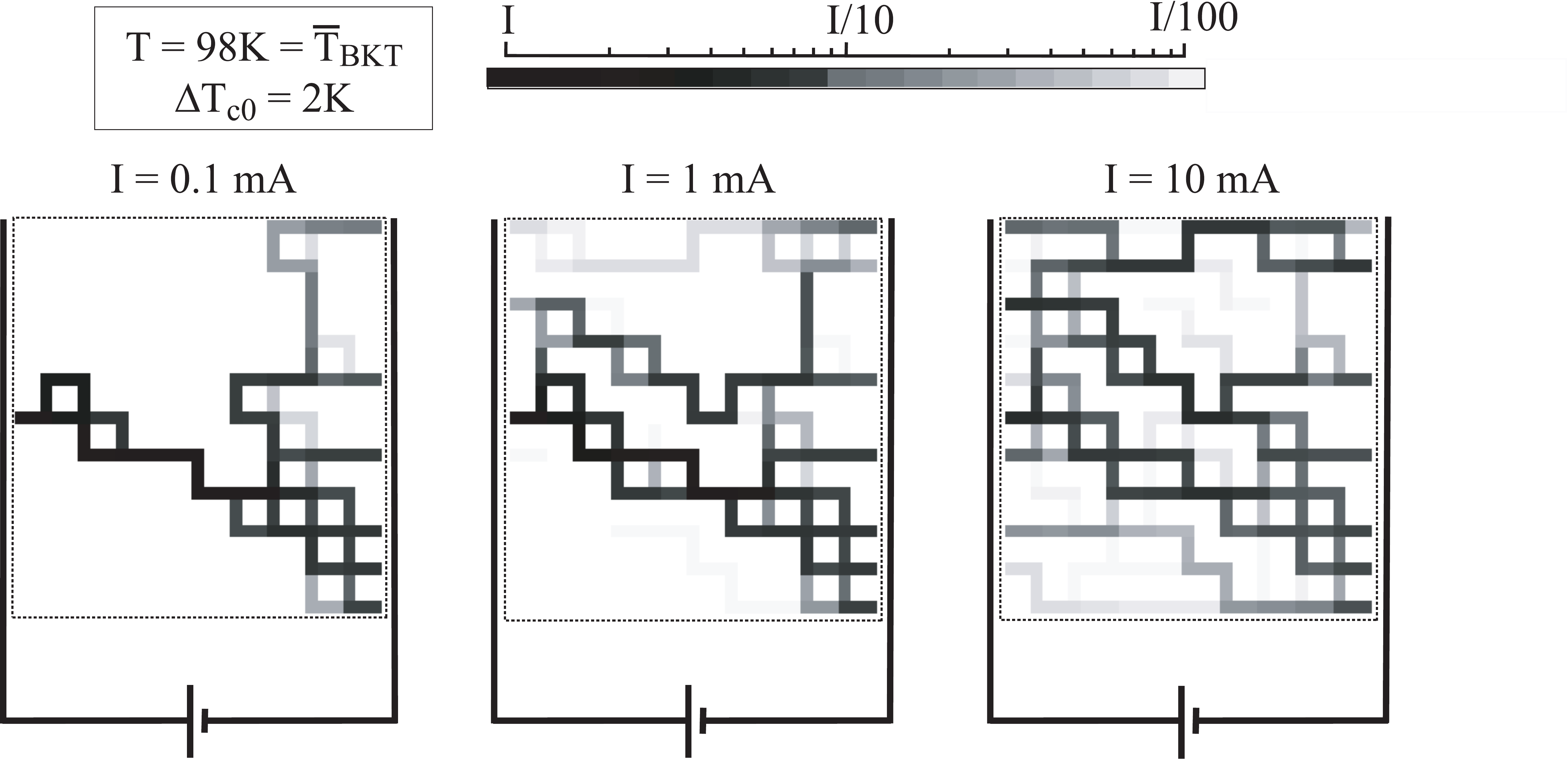}
\caption{Some examples of the current redistributions obtained in our simulations, when the bias intensity is varied and the temperature is constant. The simulation sample is the same as in \Fig{Fig1}. The fact that different regions of the inhomogeneous sample contribute to the electric transport as $I$ changes indicates that the log-log slope of $V(I)$ may vary as $I$ is varied, in spite of the power-law-like $V\propto I^\alpha$ behaviour of the homogeneous case. This change in slope is appreciable in the \Figab{Fig3}.}
\label{Fig2}
\end{figure}

Our aim is to obtain the \VI\ characteristics of a 2D superconductor composed by randomly-located domains, each domain having its own single \Tccero\ and \TBKT\ and following the \VI\ characteristics described in the previous section. For this purpose, in the spirit of the finite-element methods we model the inhomogeneous superconductor as a $N\times N$ square mesh of resistors (see \Fig{Fig1} and \Fig{Fig2}). We randomly assign to each node of the mesh a different \Tccero, and to each resistor the \Tccero\ of its corresponding left-lower node. The \Tccero\ distribution is Gaussian with mean-value \Tcceromedia\ and full-width at half-maximum \deltaTccero. The difference $\Tccero-\TBKT$ is held constant for all resistors, so that the distribution of \TBKT\ follows the one of \Tccero\ and is Gaussian with mean-value \TBKTmedia\ and full-width at half-maximum also $\deltaTccero$. We include also in our model an external circuit composed by a current source connected  to opposite borders of the sample with zero-resistance contacts. When referring to the results of our simulations, by $I$ we mean this external bias current and by $V$ the voltage drop between those opposite contacts. The sample is considered to have width and length $w$ and thickness $d$.

%%% Figure 1%%%%%

%%% Figure 2%%%%%

Because the resistivity of each resistor depends on the local current passing through it (see previous section), the mesh equations that result from that modeling are nonlinear and in general do not admit analytic solution. They have to be solved using numerical methods, of which we use Newton-type iterations. For those iterations to succeed, it is crucial to start them from initial values not too far from the solution. Thus, to calculate each \VI\ curve at a fixed temperature, we applied the following algorithm, that proved itself to be well adapted to the non-Ohmic features of the BKT transition (as it solves the instability problems that we found trying other simulation strategies): Our analysis starts by considering first a high temperature $T_{start}\gg\Tcceromedia$ (where the system is Ohmic and easy to solve) and a high bias intensity $I_{start}=(wd/10)j_0$. The analysis then evolves keeping the bias intensity constant but lowering the temperature, evaluating at each $T$-step the mesh equations by means of Newton iterations with starting point the final result of the previous step. The $T$-decrement separating each step is adaptively updated during the simulation, so that voltages are not allowed to vary beyond 0.1\% between steps. Once reached the target temperature at which the \VI\ curve is to be calculated, the temperature is fixed and then the bias intensity $I$ is varied, again adaptively, iteratively solving the mesh equations at each step and storing the results. Each $I$-step uses as starting values for the Newton iterations the results obtained in the previous calculated step. As an additional measure to avoid instabilities in the convergence of the Newton iterations, when needed our program smoothes over $0.05$K the $V(T)$ behaviour of the individual resistors at their local \TBKT\ temperature, making its $V(T)$ evolution continuous (but still very rapidly varying). This $T$-widening of the BKT transition is negligible in any case against the one due to the inhomogeneities considered in our simulations  and we checked that doubling or halving it does not change our final results in any appreciable way.

All of the results presented in this paper correspond to $10 \times 10$ cell meshes. We have checked that runs of the simulation using different random distributions but with the same statistical parameters (\Tcceromedia, \TBKTmedia\ and \deltaTccero) produce similar outcomes. Computation time for a single \VI\ curve at fixed $T$ is of about three days for a $10 \times 10$ mesh in current desktop computers.

\mbox{}

\section{Results of the simulation}

Let us now comment on the results obtained when applying the method described in the previous section to compute the \VI\ curves around the BKT transition, using parameter values typical of HTSC film samples. In particular, we have used $\Tcceromedia=100$K, $\deltaBKT=2$K (therefore $\TBKTmedia=98$K), $\sigma_n=10^{8}T^{-1} {\rm K}/\Omega{\rm m}$, $\xiabo=1\,{\rm nm}$, $d=100\,{\rm nm}$ and $w=10^{-3}{\rm m}$, and we have constructed simulated samples from $\deltaTccero=2$K to 4~K.

In \Fig{Fig1} and \Fig{Fig2} we show some examples of the current distributions within the mesh circuit, obtained for representative values of $T$ and $I$.  The most important feature observed in these results is that significant spatial redistributions of the currents may occur when either {\it T} or {\it I} are varied. This is a consequence of the fact that the resistance of each mesh element relative to the resistance of the other elements will be dependent on $T$ and $I$, and therefore the current paths will be also dependent on both of these variables.  Due to these redistributions, changes of {\it T} and {\it I} will also change what are the portions of the sample that dominate the global voltage drop, and its Ohmic or  non-Ohmic character.

We may also note that, as in fact it was already commented in the Introduction, the behaviour of the system in general will be intermediate between the simplest cases of considering all resistors in series (in which case the global resistivity is given by the average of the resistivity of all the elements) or in parallel (in which case the minimum resistance will dominate the transport properties), being the proximity to each situation dependent on $I$ and $T$. Interestingly,  for temperatures above \Tcceromedia\ the current distribution becomes especially simple: As it can be seen in \Fig{Fig1} and \Fig{Fig2}, at $T\gg\Tcceromedia$ almost all of the current flows longitudinal and uniformly. So, at those temperatures the global resistance may be approximated by the one of a  row, which in turn corresponds to the average resistance of the resistors on it. As the temperature is lowered we see however that the current path geometry is no longer that simple, nor constant with $T$ or $I$. Even for $T<\TKTmedia$ the current does not flow longitudinally, and transversal current paths remain significant. 

%%%%%%%%%%%%%%%%%%%%%%%%%%%%%%%%%%%%%%%%%%%%%%%%%%%%%%%%%%%%%%
%%%%%%%%%%%%%%%%%%%%%%%%%%%%%%%%%%%%%%%%%%%%%%%%%%%%%%%%%%%%%%

%%% Figure 3%%%%%

\begin{figure}[b!]
\begin{center}\mbox{}\vspace{1.0cm}\\
\includegraphics[width=0.47\textwidth]{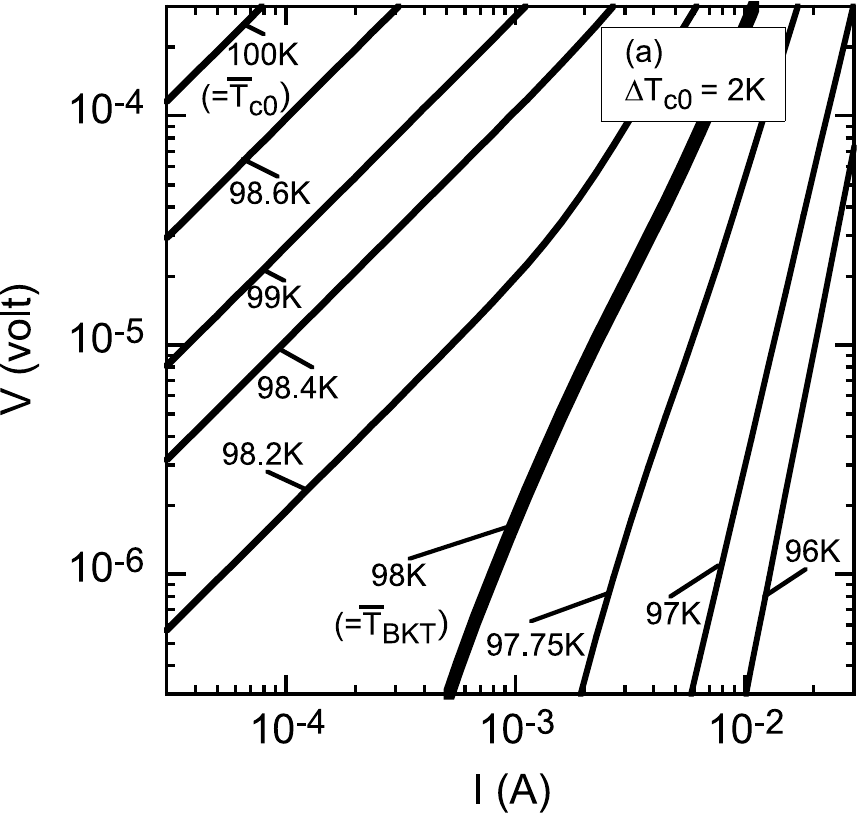}$\;\;\;\;\;\;$\includegraphics[width=0.47\textwidth]{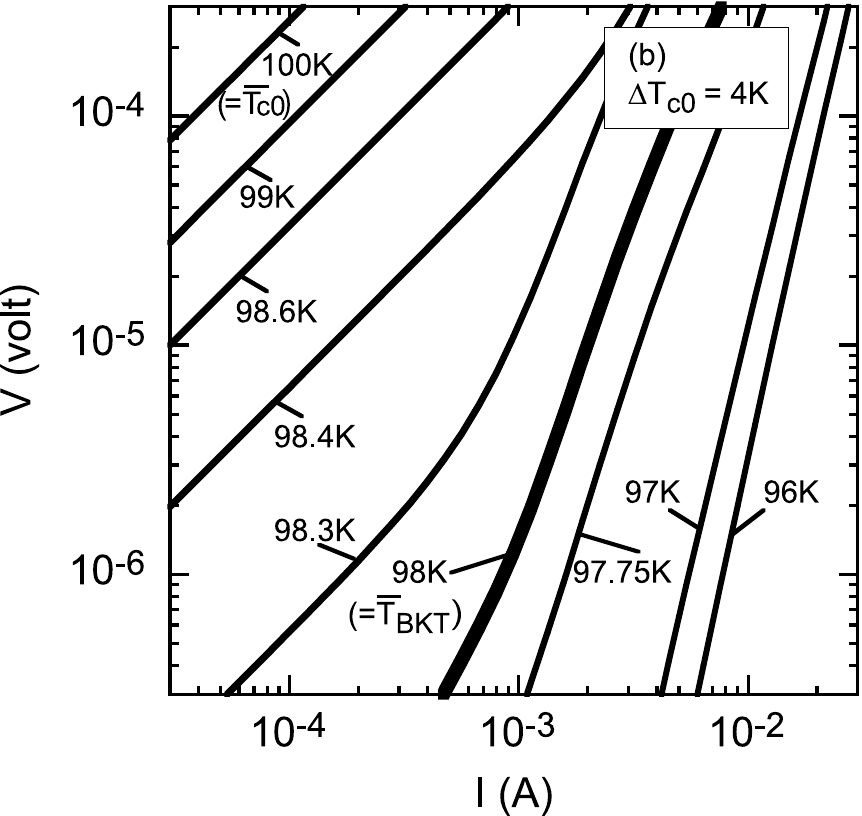}\end{center}
\caption{\VI\ results at various constant temperatures around the BKT transition as obtained in our simulations in two planar superconductors with (a) $\deltaTccero=2$K and (b) $\deltaTccero=4$K. In both cases we have used parameter values typical of HTSC films, namely $\Tcceromedia=100$K, $\TBKTmedia=98$K, $\deltaBKT=2$K, thickness 100nm, width and length $10^{-3}$m, $\sigman=10^8T^{-1}$K/$\Omega$m, and $\jiab(0)=1$nm. We have also used $b_1=1$ (see main text). The log-log slope of these \VI\ results corresponds to the exponent~$\alpha$ (see also \Fig{Fig3}). The $T=\TBKTmedia$ isotherm (thickest line) presents $\alpha=3$, while the $T=\Tcceromedia$ isotherm (upper line) is Ohmic ($\alpha=1$). Note that non-Ohmic behaviour appears in these inhomogeneous superconductors already above \TBKTmedia. Note also that the log-log slope may vary with $I$ or $V$. In this work the results reported for $\alpha(T)$ correspond to the $V$-range $3\times10^{-6}$ to $3\times10^{-5}\;\mbox{volt}$.}
\label{Fig3}
\end{figure}

%%% figure 4%%%%%
\begin{figure}[b!]
\begin{center}\mbox{}\vspace{1.5cm}\\
\includegraphics[width=0.47\textwidth]{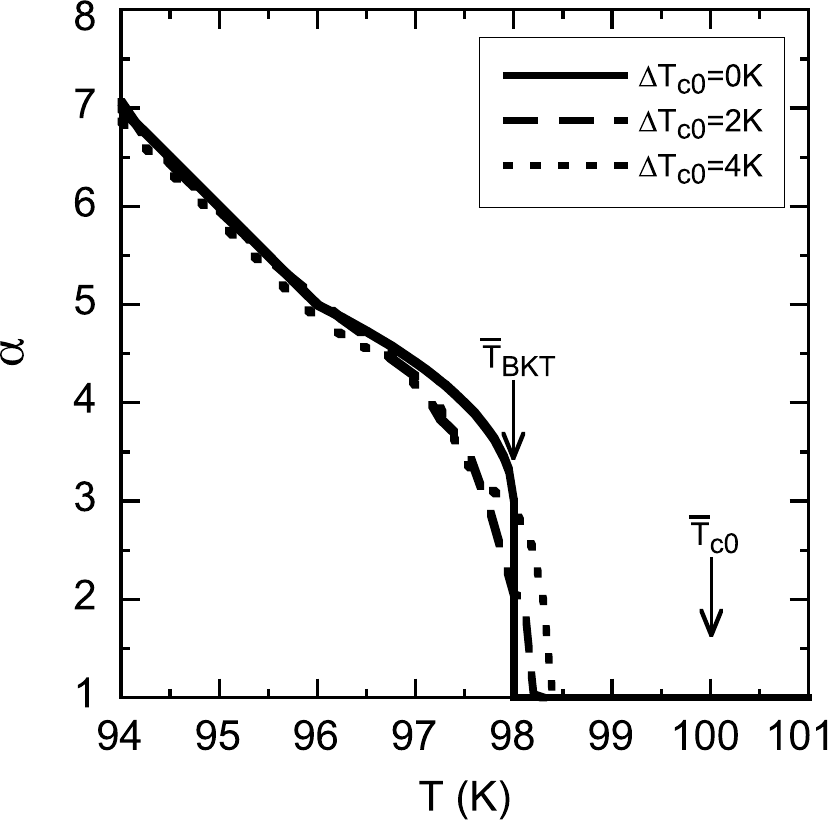}\end{center}
\caption{The exponent $\alpha$ in $V\propto I^\alpha$ extracted as the log-log slope of the simulation results shown in \Figab{Fig3}, in the range $3\times10^{-6}$ to $3\times10^{-5}{\rm volt}$. The  homogeneous case, $\deltaTccero=0$K,
 is also shown for comparison (see section 2). Note that the condition $\alpha=3$ provides a good estimate for \TBKTmedia, and that extrapolating the $\alpha(T)$ behaviour for $T<\TBKTmedia-\deltaBKT$ (=96K) to the Ohmic value, $\alpha=1$, provides a good estimate for \Tcceromedia.}
\label{Fig4}
\end{figure}

In \Figab{Fig3} we show the \VI\ curves that result from these simulations. As mentioned previously, here $V$ and $I$ correspond to the global values, \ie, those in the external bias circuit. The log-log slope of these curves corresponds to the exponent $\alpha$ in the $V\propto I^\alpha$ dependence. It is evident in \Figab{Fig3} that $\alpha$ depends on temperature. As it could be expected, Ohmic behaviour (\ie, slope unity) is obtained for temperatures well above \TBKTmedia, while much larger log-log slope is obtained well below that temperature. For temperatures close to \TBKTmedia\ the change in behaviour is not discontinuous, being instead somewhat broadened by inhomogeneities, although the $T$-range where that broadening occurs is significantly smaller than \deltaTccero. This change in the log-log slope may be seen more accurately in \Fig{Fig4}, where  $\alpha(T)$ is plotted for all of the simulated samples, together with the theoretical $\alpha(T)$ corresponding to the homogeneous case. To obtain this \Fig{Fig4},  $\alpha$ was calculated trough a power-law fit to our \VI\ results at voltages around $10^{-5}$~volt (in particular $3\times10^{-6}{\rm volt}\leq V\leq3\times10^{-5}{\rm volt}$). The reason why it is necessary to specify a voltage range for the obtainment of $\alpha$ is that, as may be noticed in \Figab{Fig3}, in the inhomogeneous samples for temperatures close to \TBKTmedia\  the \VI\ dependence is not perfectly power-like, but rather the log-log slope depends also on the applied current. The cause behind this fact is the existence of local current redistributions, shown in \Fig{Fig1} and \Fig{Fig2} (\ie, changes in $I$ vary the region of the sample where the voltage drops, and thus its global Ohmic or non-Ohmic character).

Other important feature to be observed in \Fig{Fig4} is that two of the most common criteria used  by experimentalists \cite{LTSC-1,LTSC-2,LTSC-3,Science,Nature,HTSC-1,PRB40VI,PRB42,HTSC-3,HTSC-4} to determine from the $\alpha(T)$ plots the BKT and mean-field critical temperatures remain essentially valid in spite of the inhomogeneities, if we apply those criteria to determine now their average values, \TBKTmedia\ and \Tcceromedia: In particular, the condition $\alpha=3$ provides a good estimate for \TBKTmedia, underestimating it only slightly (in particular the deviation from the exact value is well smaller than \deltaTccero). Secondly, the $\alpha(T)$ dependence for temperatures $T\Lsim\TBKTmedia-\deltaBKT$ extrapolates to the value $\alpha=1$ at $T=\Tcceromedia$ with excellent accuracy, indicating then that such extrapolation procedure remains adequate in real inhomogeneous samples to determine \Tcceromedia.

%%% Figure 5%%%%%
\begin{figure}[b!]
\begin{center}\mbox{}\vspace{1.5cm}\\
\includegraphics[width=0.47\textwidth]{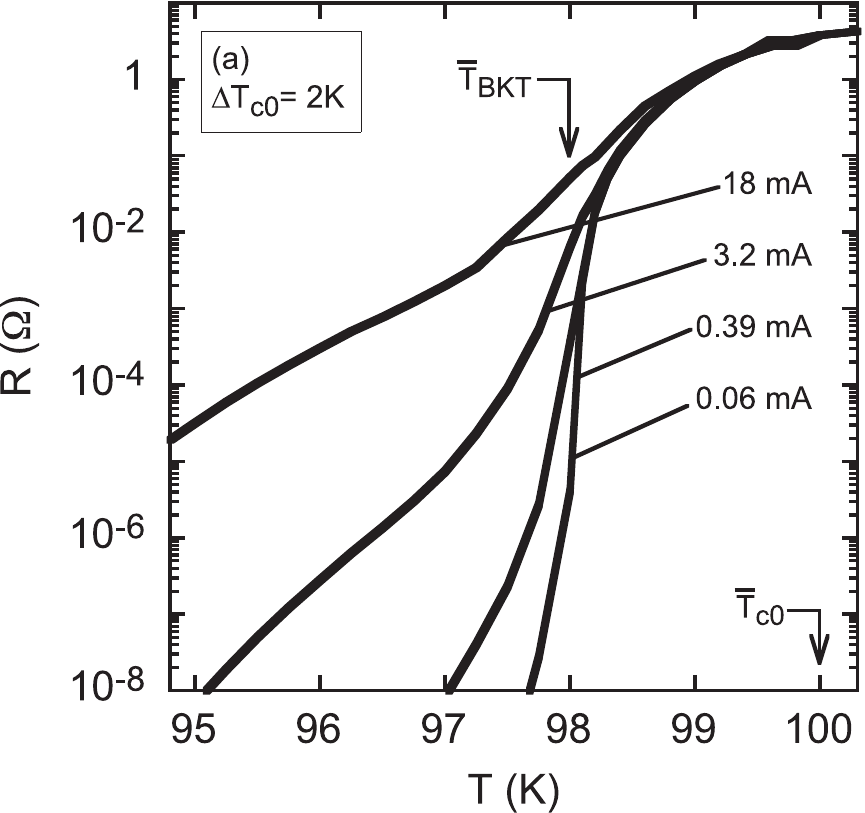}$\;\;\;\;\;\;$\includegraphics[width=0.47\textwidth]{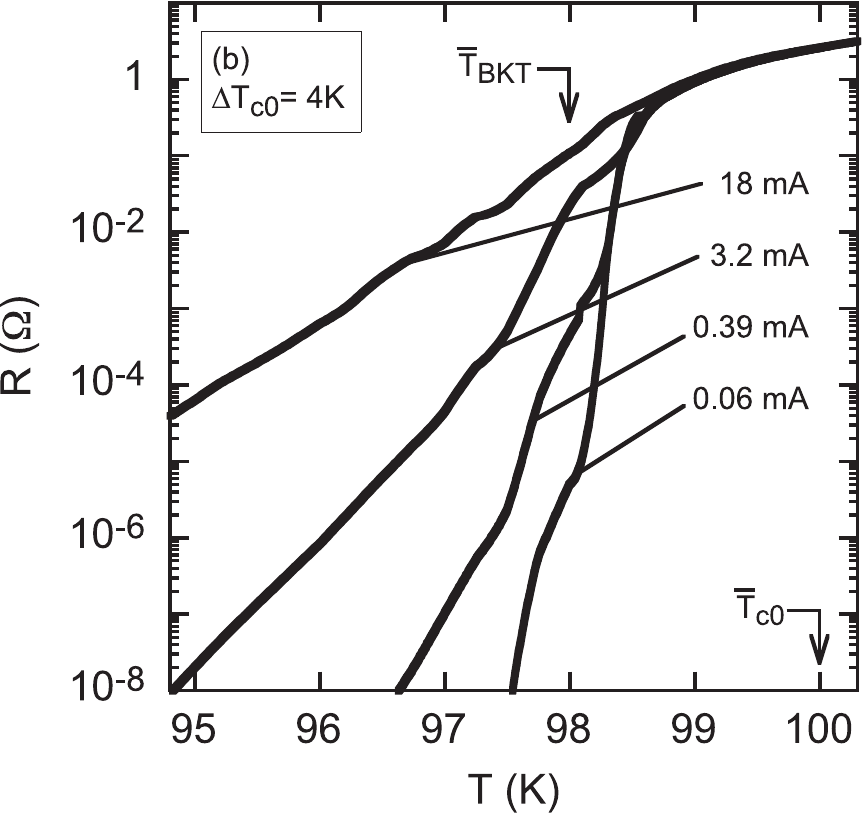}\end{center}
\caption{Resistance versus temperature at various constant bias currents as resulting from the quotient $V/I$ in the same simulation runs as in \Figab{Fig3} for superconductors with (a) $\deltaTccero=2$K and (b) $\deltaTccero=4$K. The logarithmic scale allows to better appreciate that the tail of the resistive transition is broadened when the current is increased and that the effect is larger for the samples with a larger \deltaTccero. Note also that non-Ohmic behaviour sets in already above \TBKTmedia.\vspace{12pt}\mbox{}}
\label{Fig5}
\end{figure}

The effects of the inhomogeneities in the dc electrical transport properties become more apparent in the $R(T)$ curves obtained at different fixed external currents $I$. Here we define $R$ as simply $V/I$. In \Figab{Fig5} we show the $R(T)$ results obtained from our simulations at different fixed $I$ values, for samples with different \deltaTccero\ values and using a logarithm axis for the resistance. It may be seen in these figures that the application of a finite current broadens the ${\rm log}R(T)$ tail in the lower part of the superconducting transition, and that this happens to a larger extent as \deltaTccero\ is increased. It is also easily noticeable that the non-Ohmic behaviour sets in at temperatures above \TBKTmedia.~We conclude that the inhomogeneities are detectable over a larger $T$-range on the amplitude of the resistance than on the $\alpha$ exponent.

%%%%%%%%%%%%%%%%%%%%%%%%%%%%%%%%%%%%%%%%%%%%%%%%%%%%%%%%%%%%%%%
%%%%%%%%%%%%%%%%%%%%%%%%%%%%%%%%%%%%%%%%%%%%%%%%%%%%%%%%%%%%%%%

\begin{figure}[b!]
\begin{center}\mbox{}\vspace{1.5cm}\\
\includegraphics[width=0.47\textwidth]{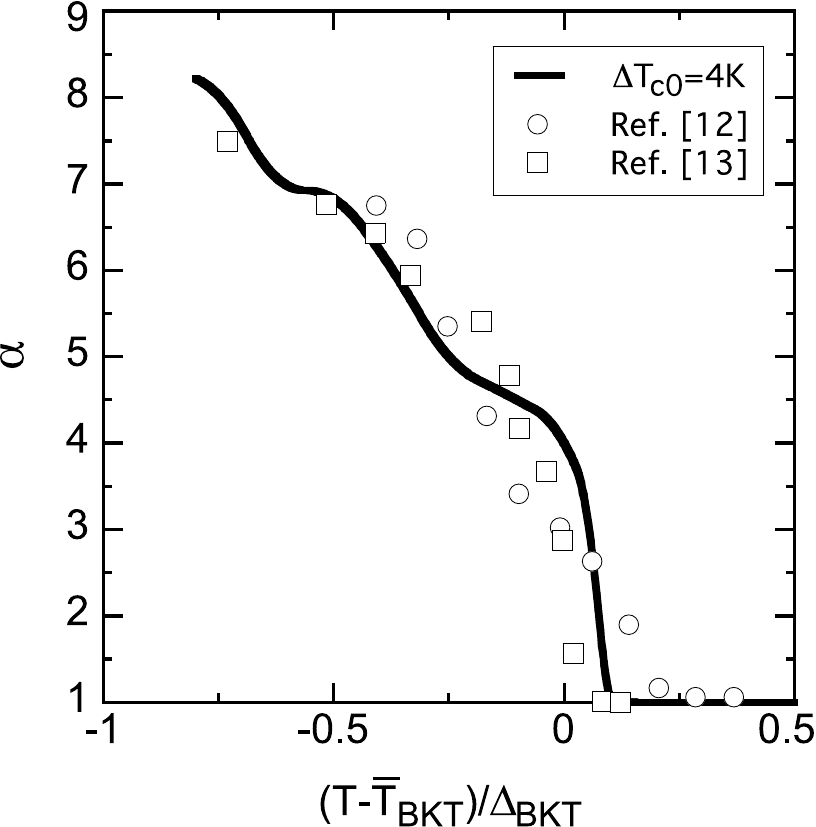}\end{center}
\caption{Comparison between the exponent $\alpha$ obtained experimentally in \Refs{PRB40VI} and \cite{PRB42} in HTSC (circles and squares) and our simulations (solid line). The experimental data were obtained by means of \VI\ measurements in \TlBCCO\ 700nm-thick films \cite{PRB40VI} and in \YBCO\ 120nm-thick films \cite{PRB42}. In this comparison we employed for \TBKTmedia\ the values that result from the condition $\alpha=3$, and for \Tcceromedia\ the values that result from extrapolating to $\alpha\rightarrow 1$ the low-temperature $\alpha(T)$ data. We obtained $\TBKTmedia=99.0$K and $\Tcceromedia=100.2$K for \Ref{PRB40VI} and $\TBKTmedia=83.45$K and $\Tcceromedia=85.95$K for \Ref{PRB42}. We also used the values $\deltaTccero=4$K and $b_1=2$, that produce the best agreement with these $\alpha(T)$ data.}
\label{Fig6}
\end{figure}

Let us now discuss how the results obtained with our simulation procedure compare with some of the experimental data obtained by earlier authors measuring the \VI\ characteristics near the BKT transition. Specifically, we will use for these comparisons the $\alpha(T)$ measurements of \Refs{PRB40VI} and \cite{PRB42}, as in these works $\alpha(T)$ was extracted using a voltage criterion similar to the one used in our simulations. These measurements were performed in \Ref{PRB40VI} in \TlBCCO\ 700nm-thick films (see  in particular figure~3(a) of that work) and in \Ref{PRB42} in \YBCO\ 120nm-thick films (see figure~2 of that work). We show in our \Fig{Fig6} a comparison between these experimental $\alpha(T)$ and our simulation results. To be able to gather together in a single representation the two samples in spite of their different critical temperatures, we have chosen for the horizontal axis the normalized quantity $(T-\TBKTmedia)/\deltaBKT$. We employed for each sample the \TBKTmedia\ that results from applying the condition $\alpha(\TBKTmedia)=3$, and the \Tcceromedia\ that results from extrapolating to $\alpha\rightarrow 1$ the low-temperature $\alpha(T)$ data. We obtained $\TBKTmedia=99.0$K and $\Tcceromedia=100.2$K for \Ref{PRB40VI} and $\TBKTmedia=83.45$K and $\Tcceromedia=85.95$K for \Ref{PRB42}. In our simulations we have used the same parameter values as for the simulations shown in figures 1 to 5, except for $\deltaTccero=4$K and $b_1=2$. These latter values were found to be the ones producing a better agreement between those data and our simulations (note that in the case of $b_1$ its value is governed mainly by the data for $T<\TBKTmedia-\deltaBKT$, outside of the region significantly affected by inhomogeneities). As may be seen in the \Fig{Fig6}, the agreement between experiments and simulation is rather satisfactory, despite the necessary crudeness of some of our approximations (perfectly Gaussian distribution of the \Tccero-inhomogeneities, uniform \deltaBKT\ value, inexactness of any finite-element method, etc.).

%%% Figure 6%%%%%

%%%%%%%%%%%%%%%%%%%%%%%%%%%%%%%%%%%%%%%%%%%%%%%%%%%%%%%%%%%%%%%
%%%%%%%%%%%%%%%%%%%%%%%%%%%%%%%%%%%%%%%%%%%%%%%%%%%%%%%%%%%%%%%

\mbox{}

\section{Conclusions}

We have analyzed numerically the effects of a random spatial Gaussian distribution of critical temperature inhomogeneities with long characteristic lengths on the voltage-current \VI\ characteristics of a type-II planar superconductor near the Berezinskii-Kosterlitz-Thouless (BKT) transition. The simulations allow to quantify the broadening around the average BKT transition temperature \TBKTmedia\ of both the exponent $\alpha$ in $V\propto I^{\alpha}$ and of the resistance $V/I$. These calculations reveal that strong spatial redistributions of the local current will occur around the transition as either $I$ or $T$ are varied. Our results also support that the condition $\alpha=3$ provides a good estimate for the location of the average BKT transition temperature \TBKTmedia, and that extrapolating to $\alpha\rightarrow 1$ the $\alpha(T)$ behaviour well below the transition provides a good estimate for the average mean-field critical temperature \Tcceromedia. These results are in good agreement with some experimental measurements of the exponent $\alpha(T)$ obtained by earlier authors on HTSC films.\cite{PRB40VI,PRB42}

\mbox{}

\mbox{}\\ {\Large \bf Acknowledgements}\\ \mbox{}\\
\nonfrenchspacing
N.~Cot\'on acknowledges financial support from Spain's Ministerio de Ciencia e Innovaci\'on (MICINN) under project FIS2007-63709 (MEC-FEDER) trough a FPI grant. This work has been also supported by the MICINN project FIS2010-19807 and by the Xunta de Galicia projects  2010/XA043 and 10TMT206012PR. All these projects are co-funded by ERDF from the European Union.

\newpage

\end{document}